%% file: TDGI_Preprint.tex
\documentclass[preprint,showpacs,preprintnumbers,amsmath,amssymb]{revtex4-1}
\usepackage{graphicx}

\begin{document}

\title{Differential ghost imaging in time domain}

\author{Yoshiki O-oka}
\affiliation{Graduate School of Arts and Sciences, University of Tokyo, Komaba, Meguro, Tokyo 153-8902, Japan}

\author{Susumu Fukatsu}
\email{cfkatz@mail.ecc.u-tokyo.ac.jp}
\affiliation{Graduate School of Arts and Sciences, University of Tokyo, Komaba, Meguro, Tokyo 153-8902, Japan}

\begin{abstract}
Differential ghost imaging was attempted in time domain, i.e., temporal differential ghost imaging (TDGI), using pseudo-randomized light pulses and a temporal object consisting of no-return-to-zero bit patterns of varying duty. Evaluation of the signal-to-noise characteristics by taking account of errors due to false cross-correlation between the reference and the bucket detector readings indicates that the TDGI outperforms its non-differential counterpart, i.e., time-domain GI, in terms of consistently high and even duty-independent signal-to-noise ratios that are achieved. Dynamic local averaging helps save data recording without compromising the essential features of the TDGI. 
\end{abstract}


\maketitle
\newpage

Ghost imaging (GI) is a technical framework that allows image retrieval by sharing private keys between two parties\cite{pittman1995optical}. In a typical GI setup, an object to be imaged is placed in one of two arms of light passage with a bucket detector which provides only the timing information of photon arrival at the expense of spatial resolution. The other \textit{empty} arm is solely equipped with a space-resolving detector, from which no information of the object is available. Correlating the set of \textit{sparse} data acquired in the two arms, each of which in itself carriers no meaningful information thereby forming a private key, can only reconstruct the image of the object that is otherwise unretrievable.\cite{pittman1995optical, bennink2002two, valencia2005two, ferri2005high,  shapiro2008computational, katz2009compressive, meyers2011turbulence, clemente2010optical, sun20133d, gao2017distributed} Because of its inherent cryptic nature of the information retrieval scheme and the simplicity of implementation, the GI can offer a rich variety of applications not limited to 2-D imaging\cite{liu2007fourier, hozawa2013single, shi2014polarimetric, jha2015spectral}. Although GI was exclusively discussed in space domain, a recent report has pointed out that the lens-less GI concept can be potentially transplanted to time domain\cite{ryczkowski2016ghost}. Such a time-domain analogue of the GI and its allies seem to hold promise in light of metrology and information processing where time varying signal plays a pivotal role, as opposed to only slowly varying, if not static, spatial information\cite{shirai2010temporal, chen2013temporal, devaux2016computational, ryczkowski2016ghost, devaux2016temporal, ryczkowski2017magnified}. 
	
Signal-to-noise characteristics are an issue of significance where information retrieval is relevant. In this regard, however, the original GI is not always best placed, and variants of GI have been developed to circumvent limitations arising from otherwise inadequate signal-to-noise ratio (SNR). Differential GI (DGI) is one such scheme that holds promise in achieving an improved SNR \cite{ferri2010differential, wang2015gerchberg}. In this study, an attempt is made to implement the DGI in time domain, i.e., temporal DGI (TDGI), which is discussed from the data-processing point of view.
	
\begin{figure}[!t]
\begin{center}
\includegraphics[width =0.85 \linewidth]{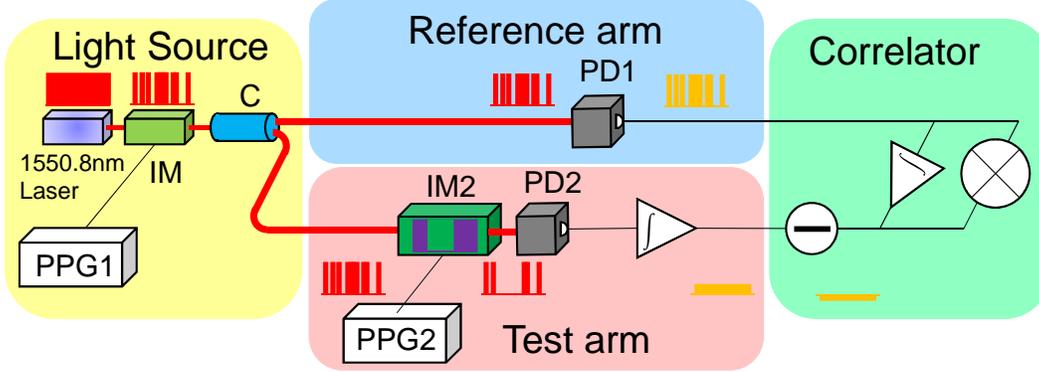}
 \end{center}
 \caption{Setup of time-domain differential ghost imaging (TDGI). Time-bucket detection is accomplished by integrating time-resolved signal on the test arm, $I_{\rm{test}}$ transmitted through the intensity-only temporal mask. IM1(2): intensity modulator, PPG1: pulse pattern generator (2$^{11}-1$ PRBS, 100 Mbps) C: 3 dB coupler, PPG2: 12.5-Mbps pulse pattern generator, PD1(2): photodiode, O: oscilloscope}
\label{Fig.1}
\end{figure}

Similarly to the DGI in space domain, the TDGI is based upon measurements of the cross-correlation between two arms, one of which is equipped with an object that time-encodes the information to be shared by joint detection. All experiment was done by using fiber optics (Fig. \ref{Fig.1}). The light source was a telecom-grade laser tuned at 1550.8 nm (Santec TSL-210), the output of which was attenuated and coupled to an intensity modulator (JDSU 11020416). Chaotic light pulses of 100-\% modulation depth with speckle-like distribution were mimicked by driving the intensity modulator with a pulse pattern generator (Agilent 8110A) producing 100 Mbps non-return-to-zero (NRZ) pulse streams according to a $2^{11}-1$ pseudo-randomized bit sequence (PRBS). A 3-dB coupler split the input light into two beams along the test and the reference arm. The latter was coupled to a 5-GHz time-resolving detector (Thorlabs DET08CFC), whereas the former was equipped with an intensity modulator as the programmable intensity-only temporal object, i.e., time mask, and a \textit{time} bucket detector emulated by integrating the time-resolved signal from a second 5-GHz wide-band detector such that $B=\int^T_0{dt_{2}}I_{\rm{test}}(t_{2})$ where $I_{\rm{test}}$ is the detector reading in the reference arm with $T$ being the detection period.

The TDGI is based on the covariance 
\begin{eqnarray}
C_{\rm TDGI}\left( t_1 \right)=\langle\mathit \Delta I_{\rm ref}\left( t _{1} \right) \mathit{\Delta} B_{-}\rangle_{N}\label{C}
\end{eqnarray}
over an ensemble of $N$ systems where $\mathit{\Delta}$ represents the fluctuation, $I_{\rm{ref}}$ is the detector reading in the reference arm, and

\begin{equation}
B_{-}=B-\overline{T}\int{dt_{1}}I_{\rm{ref}}(t_{1})
\label{B_}
\end{equation}
is the \textit{differential} time-bucket detector reading with $\overline{T}={\left<B\right>}/{\left<\int{dt_{1}}I_{\rm{ref}}(t_{1})\right>}$ being the mean transmittance of the temporal object seen in the reference arm. 

Rewriting Eq. (\ref{C}), we find
\begin{eqnarray}
\nonumber
C_{\rm TDGI}\left( t_1 \right)&=&\int dt_{2}\langle\mathit{\Delta} I_{\rm{ref}}(t_{1})\mathit{\delta} T(t_2) \mathit{\Delta} I_{\rm{test}}(t_{2})\rangle_N\\
\nonumber
&=&\langle\mathit{\Delta} I_{\rm{ref}}(t_{1})\mathit{\delta} T(t_{2}=t_{1}) \mathit{\Delta} I_{\rm{test}}(t_{2}=t_{1})\rangle_N\\
&&+\int_{t_1\ne t_2}dt_{2}\langle\mathit{\Delta} I_{\rm{ref}}(t_{1})\mathit{\delta} T \mathit{\Delta} I_{\rm{test}}(t_{2})\rangle_N
\end{eqnarray}
where $\mathit{\delta}T \equiv T(t_2)-\overline{T}$. The first term on right-hand side is the \textit{signal} due to the legitimate autocorrelation 
\begin{equation}
({\rm Signal})\simeq \langle\mathit{\Delta} I_{\rm{ref}}(t_{1})\rangle^2_N\mathit{\delta} T(t_{1})\\
\end{equation}
since $\mathit{\Delta} I_{\rm{ref}}(t_{1})=\mathit{\Delta} I_{\rm{test}}(t_{1})$, whereas the second term is the \textit{error} due to false cross-correlations
\begin{equation}
({\rm Error})\simeq
\langle\mathit{\Delta} I_{\rm{ref}}(t_{1})\rangle_N \overline{\mathit{\delta} T} \langle\mathit{\Delta} I_{\rm{test}}(t_{2})\rangle_N
\end{equation}
where $\overline{\mathit{\delta} T}=\int dt_{2}\mathit{\delta} T(t_2)/\int 1 dt_{2}$. Such errors are inevitable for light pulses of speckle-like distribution, which can compromise the integrity of images to be retrieved.

\begin{figure}[!b]
\begin{center}
\includegraphics[width = 0.85\linewidth]{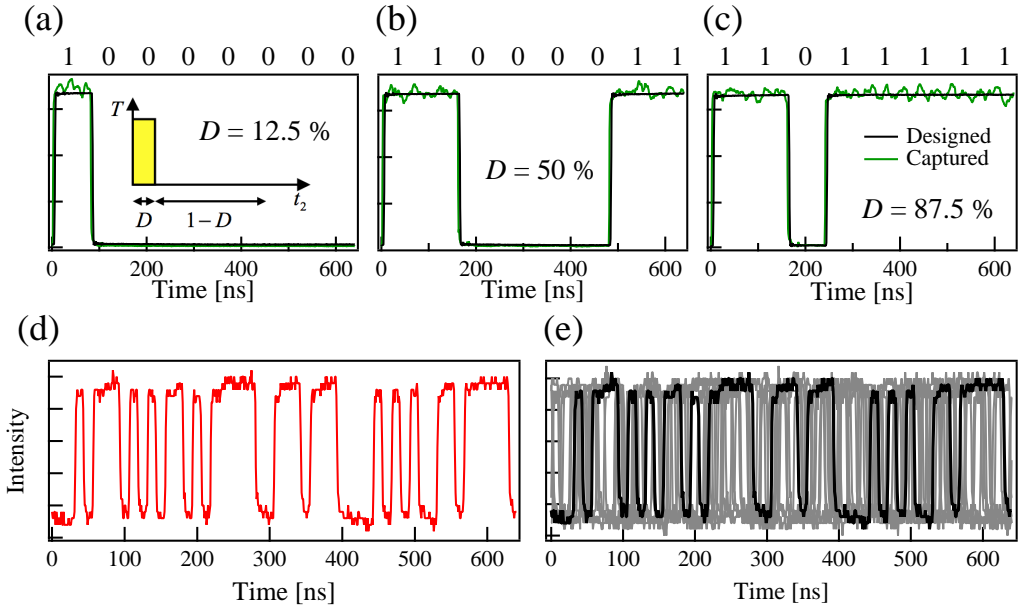}
\end{center}
\caption{Oscilloscope traces (green) of the time mask patterns for three-bit NRZ pulses of varying duty, $D$, generated by a pulse pattern generator (black): 12.5\% (a), 50\% (b), and 87.5\% (c). Figures above the top axis indicates the binary data. The inset illustrates how the mask duty $D$ relates to $T$, from which $\overline{T^2}=D$ and $\overline{\mathit \delta T^2}=D(1-D)$ are found as described in the text. The lower insets show a snap shot of the input PRBS pulse (c) and its cumulative recordings (d) that ensure uniform temporal distribution.}
\label{Fig.2}
\end{figure}

The SNR is defined as

\begin{equation}
\left(\rm{SNR}\right)_{\rm{TDGI}} =\frac{\overline{\langle C_{\rm TDGI}\rangle ^{2}}}{\overline{\left<\left[C_{\rm TDGI}(t_1)-\langle C_{\rm TDGI}\right>\right]^2\rangle}}.
\label{SNR_TDGI_0}
\end{equation}
where $\overline{\langle A \rangle}=\int dt_{1}\langle A(t_1)\rangle_M/\int dt_{1}$ with $M$ being the repetition number of measurements. 

Figures \ref{Fig.2}(a)-(c) show three-bit NRZ time mask patterns with varying duty, $D$, of the time mask ( (a) 12.5 \%, (c) 50 \%, (c) 87.5 \%): as-designed (black) and captured (green).  Fig. \ref{Fig.2}(d) shows the typical profile of a 2$^{11}$-1 PRBS pulse. Uniform distribution of pulses is established by accumulating many different pulse patterns (Fig. \ref{Fig.2}(e)). Assuming a flat-top distribution of $\langle I_{\rm{ref}} \rangle$ over a large ensemble, Eq. (\ref{SNR_TDGI_0}) reduces to\cite{o-oka1}

\begin{equation}
\left(\rm{SNR}\right)_{\rm{TDGI}} \propto \frac{\overline{\mathit{\delta} T^{2}}}{\overline{\mathit{\delta} T^{2}}}.
\label{SNR_TDGI_2}
\end{equation}

\begin{figure}[!b]
\begin{center}
\includegraphics[width = 0.8\linewidth]{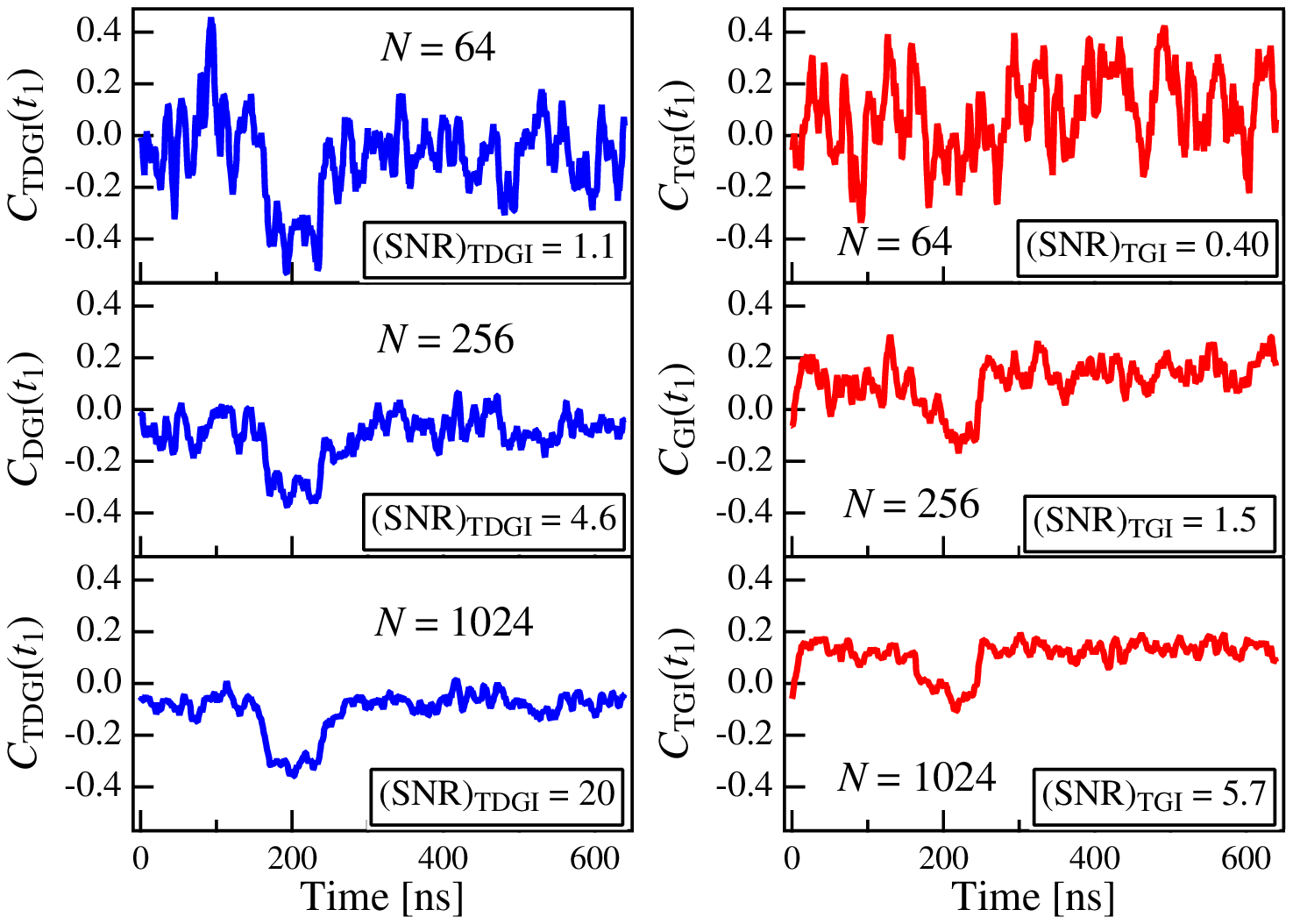}
\end{center}
\caption{$C_{\rm TDGI}\left(t_1\right)$ (left column) and $C_{\rm TGI}\left(t_1\right)$ (right column) versus$t_1$. Figures in the boxes are the SNR values. Three panels in each column demonstrate the effect of averaging over an ensemble of $N$. Note the "dip" in the mask profile which is clearly visible for the TDGI even at an SNR of $\approx$ 1. }
\label{Fig.3}
\end{figure}

For the time-domain version of normal GI, viz. temporal GI (TGI), however, $B$ replaces $B_{-}$ in Eq. (\ref{C}) to give $C_{\rm TGI}\left( t_1 \right)$. Accordingly, a difference from the $t_1$-average, $C_{\rm TGI}-\overline{C_{\rm TGI}}$, is relevant instead, so we have that\cite{o-oka1}

\begin{equation}
\left(\rm{SNR}\right)_{\rm{TGI}} \propto \frac{\overline{\mathit{\delta} T^{2}}}{\overline{\mathit T^{2}}}
\end{equation}
where $\overline{T}=\int dt_{2}T(t_2)/\int 1 dt_{2}$.

The inset of Fig. \ref{Fig.2} shows that $\overline{T}=D\times1+(1-D)\times0=D$ and $\overline{{T}^2}=D\times1^2+(1-D)\times0^2=D$. Therefore it follows that $\overline{\delta T^2}=\overline{\left(T-\overline{T} \right)^2}=(1-D)^2D+(-D)^2(1-D)=D(1-D)$. As such, $\left(\rm{SNR}\right)_{\rm{TGI}}$ steadily decreases as $D$ increases because $\overline{\delta T^2}/\overline{T^2}=D(1-D)/D=1-D$, whereas $\left(\rm{SNR}\right)_{\rm{TDGI}}$ is essentially independent of $D$ since $\overline{\delta T^2}/\overline{\delta T^2}=1$. From such characteristics, one can conclude that the TDGI outperforms the TGI in terms of SNR figures regardless of $D$, and that the TDGI's advantages become pronounced at high transmission, $D\approx 1$. The discussion so far is valid only if prior information on the object is available. 

Without prior knowledge of the time mask, however, the SNR values must be calculated by taking an average over the \textit{acquired} data. Using the relations $\delta T\left(t_1\right)=\langle \delta O_N \left(t_1\right)\rangle$ for the TDGI and $\delta T\left(t_1\right)=\langle O_N\left(t_1\right)-\overline{O_N\left(t_1\right)} \rangle$ for the TGI, one finds that\cite{o-oka1}
\begin{equation}
\left(\rm{SNR}\right)_{\rm{TDGI}} =\frac{\overline{\langle \mathit\delta O_N(t_1)\rangle ^{2}}}{\overline{\left<\left[\mathit\delta O_N(t_1)-\langle \mathit\delta O_N(t_1)\right>\right]^2\rangle}}
\label{SNR_TDGI}
\end{equation}
and 
\begin{equation}
\left(\rm{SNR}\right)_{\rm{TGI}} =\frac{\overline{\langle O_N(t_1)- \overline{O_N(t_1)}\rangle^{2}}}{\overline{\left<\left[O_N(t_1)-\overline{O_N(t_1)}-\langle O_N(t_1)-\overline{O_N(t_1)} \rangle\right]^2\right>}}
\label{SNR_TGI}
\end{equation}
where $O$ represents the \textit{measured} transmittance, $N$ is of the same meaning as in Eq. (1). If the repetition number, $M$, is taken to be large enough, i.e., $M \gg N$, Eqs. (9) and (10) will not essentially depend on $M$.

\begin{figure}[!b]
\begin{center}
\includegraphics[width = 0.75\linewidth]{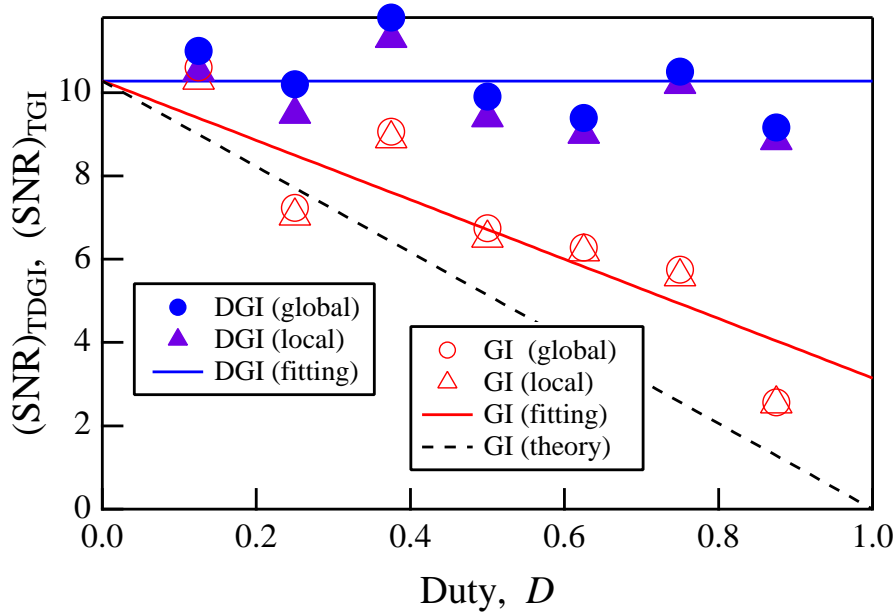}
\end{center}
\caption{Signal-to-noise ratio (SNR) as a function of the duty, $D$: TDGI (red) and TGI (blue). Circles represent the experimental data and solid lines are due to their least-squares fitting. Broken line is after theory for TGI ($\propto 1-D$). Triangles show the data obtained by \textit{local} averaging, which should be compared with circles calculated by using \textit{global} averaging.}
\label{Fig4}
\end{figure}

Figure \ref{Fig.3} compares $C_{\rm TDGI}\left(t_1\right)$ (left column) and $C_{\rm TGI}\left(t_1\right)$ (right column) plotted as a function of time, $t_1$, in the reference arm. The time mask was encoded as $"11011111" (D=0.875)$. The three panels for each column show the effect of averaging over an ensemble of varying size, $N$. The SNR values are found to scale with $N$, as opposed to $\sqrt N$, as expected from the definition of the SNR's (Eqs. (\ref{SNR_TDGI}) and (\ref{SNR_TGI})) based on quadratic estimates. Apparently, $\left(\rm{SNR}\right)_{\rm{TDGI}}>\left(\rm{SNR} \right)_{\rm{TGI}}$ holds for a given $N$ with their ratios amounting to 2.7-3.5, which arguably verifies the advantage of the TDGI over the TGI. Interestingly, the "dip" due to "0" in the time mask profile is more clearly visible in the TDGI than in the TGI, which becomes evident with decreasing SNR values. Such observations are of phenomenological significance in the sense that SNR values $\approx 1$ do not necessarily compromise the image visibility although much need to be clarified. 

Figure \ref{Fig4} is a plot of SNR values as a function of $D$. The circles are from the experiment while the solid lines are the results of least-squares fitting. As expected, the TDGI is largely independent of $D$, which is consistent with Eq. (\ref{SNR_TDGI_2}), i.e., $\left(\rm{SNR}\right)_{\rm{TDGI}}\propto \overline{\mathit{\delta} T^{2}}/\overline{\mathit{\delta} T^{2}}=1$. On the other hand, the TGI data fall almost linearly with $D$, which seems to be in qualitative agreement with the theoretical prediction $\left(\rm{SNR}\right)_{\rm{TGI}}\propto1-D$. However, the correct one, shown by the broken line, is lying even lower than the fitting trace.  Such a discrepancy is not at all unexpected and is attributed to large signal ($\sigma$) and noise  ($\nu$) values as $D\rightarrow 1$. This is because more light can provably push otherwise small SNR values upward when the intended SNR value is less than $(\sigma / \nu)^2$.

So far, we have used the \textit{full} data set of variables $\{x_i\}$ and $\{y_i\}$ in calculating the variance $C\left(x,x\right)_{N}=\sum_{i=1}^{N} (\mathit \Delta x_i)^2$ and the covariance $C\left(x,y\right)_{N}=\sum_{i=1}^{N} \mathit \Delta x_i \mathit \Delta y_i$ for an ensemble of $N$, which requires as much memory. Instead of such a trivial choice of \textit{global} averaging that is taken only after measurements are over, a dynamic \textit{local} alternative is conceivable that regularly updates a real-time local average, $m_{n}=\frac{1}{n}\sum_{i=1}^{n}x_{i}\ (1\le n \le N)$. This is favorable since it saves data by a factor of $N$ and thus consumes less memory in computing. To this end, an iterative method is useful by noting that
\begin{equation}
\mathit{\Delta} x_n =x_n - m_n \ne x_n - \frac{1}{N} \sum_{i=1}^N x_i.
\end{equation}

Compared in Fig. \ref{Fig4} are the influences of averaging: dynamic local (triangles) and full-set global (circles). For the former, it is inherently difficult to obtain as correct an average value as necessary for small $n$ values solely due to the lack of data. On the other hand, the global averaging consistently provides a more accurate and higher SNR value, which is visible in Fig. \ref{Fig4}. Nevertheless, the discrepancy remains to within 1-7 \%, which is acceptable for many purposes. Thus it is inferred that dynamic local averaging can effectively replace the global averaging by saving data recording at least and even processing time. 

In summary, differential ghost imaging was attempted in time domain using a chaotic light source and a temporal mask of a 2$^{11}$-1 PRBS pattern with varying duty. The SNR was evaluated by considering accidental errors due to false cross-correlations between the reference and time-bucket signals. The TDGI was consistently better than the TGI in terms of SNR for any duty. The developed 100-MHz technology can be easily extended by component upgrades, so wide-band operation is not an obstacle.	
	
Technical assistance from Y. Yasutake is gratefully acknowledged. This work was in part supported by JSPS KAKENHI 16K13714.

\input{TDGI_Preprint.bbl}

\end{document}

%% file: TDGI_Preprint.bbl
%